\documentclass[
aps,prb,reprint,superscriptaddress,floatfix
]{revtex4-2}

\usepackage{graphicx}
\usepackage{bm}
\usepackage{comment}
\usepackage{amssymb}
\usepackage{hyperref}
\usepackage{orcidlink}
\usepackage{amsmath}
\usepackage{bm}
\usepackage{bbm}

\begin{document}

\title{Variational Scarring in Open Two-Dimensional Quantum Dots}

\author{Fartash Chalangari \orcidlink{0009-0001-7182-204X}}
\affiliation{Computational Physics Laboratory, Tampere University, P.O. Box 600, FI-33014 Tampere, Finland}
\author{Joonas Keski-Rahkonen \orcidlink{0000-0002-7906-4407}}
\affiliation{Computational Physics Laboratory, Tampere University, P.O. Box 600, FI-33014 Tampere, Finland}
\affiliation{Department of Physics, Harvard University, Cambridge, Massachusetts 02138, USA}
\affiliation{Department of Chemistry and Chemical Biology, Harvard University, Cambridge, Massachusetts 02138, USA}
\author{Simo Selinummi \orcidlink{0009-0007-6542-6562}}
\affiliation{Computational Physics Laboratory, Tampere University, P.O. Box 600, FI-33014 Tampere, Finland}
\author{Esa Räsänen \orcidlink{0000-0001-8736-4496}}
\affiliation{Computational Physics Laboratory, Tampere University, P.O. Box 600, FI-33014 Tampere, Finland}
\date{\today}

\begin{abstract}
Quantum scars have recently been directly visualized in graphene quantum dots (Nature \textbf{635}, 841 (2024)), revealing their resilience and influence on electron dynamics in mesoscopic systems. Here, we examine variational scarring in two-dimensional quantum dots and demonstrate that these states remain robust even in an open system. We show that controlled perturbations enable modulation of electronic transmission via scarred states, presenting a viable approach to tuning quantum transport. These findings provide insights into the role of scarring in mesoscopic transport and open pathways for experimental realization in quantum devices.
\end{abstract}

\maketitle

\section{Introduction}
\noindent
Quantum chaos -- sometimes referred to as quantum chaology~\cite{Berry_bakerian_lecture_1987,Berry_1989} -- explores the transition from quantum states to classical behavior, a topic of intense debate since the advent of quantum mechanics~\cite{einstein_verh.dtsch.phys.ges_19_82_1917, stone_phys.today_58_37_2005}. While classical mechanics exhibits true chaos, quantum systems display signatures of chaotic behavior in distinct ways (see, e.g., Ref.~\cite{Stockmann_book} for a general introduction), such as characteristic spectral statistics~\cite{Gutzwiller_book}, wavefunction scarring~\cite{Heller_book}, and other non-ergodic phenomena~\cite{Casati2022_quantumChaos}. At the heart of this exploration lies the correspondence principle, which is intimately linked to longstanding questions regarding measurement and decoherence in quantum mechanics~\cite{decoherence_chaos_PhysRevLett.80.4361, Joos1985TheEO}. One proposed framework for understanding the emergence of classicality is quantum Darwinism~\cite{Zurek_rev.mod.phys_75_715_2003, Zurek_physics.today_68_9_2015, Darwinism}, which posits that decoherence -- by suppressing most quantum superpositions -- yields environment-induced superselection (\emph{einselection}). This process favors the survival of robust, \emph{pointer states} that proliferate their information into the environment, while fragile quantum states are suppressed~\cite{Darwinism}. Intriguingly, these pointer states often exhibit quantum scarring, thus defying the expected ergodic behavior of quantum-chaotic systems~\cite{Gutzwiller_j.math.phys_12_343_1971}.

Among the most striking manifestations of quantum chaos are scars~\cite{Heller_phys.rev.lett_53_1515_1984}, where certain eigenstates exhibit enhanced probability density along moderately unstable classical periodic orbits (POs). On the classical side, there is no direct counterpart for quantum scars which represent non-ergodic states allowed by the quantum ergodicity theorems~\cite{Colindeverdiere_comm.math.phys_102_497_1985, zelditch_duke.math.j_55_919_1987, Shnirelman_Uspekhi.Mat.Nauk_29_181_1974}. Quantum scars have been supported by vast theoretical evidence~\cite{kaplan_ann.phys_264_171_1998, Kaplan_nonlinearity_12_R1_1999, bogomolny_physica.d_31_169_1988, berry_proc_r_soc_lond_a_423_219_1989, kus_phys.rev.a_43_4244_1991, dariano_phys.rev.a_45_3646_1992, tomsovic_phys.rev.lett_70_1405_1993, revuelta_phys.rev.e_102_042210_2020,agam_j.phys.a.math_26_2113_1993,bohigas_phys.rep_223_43_1993, wisniacki_phys.rev.lett_97_094101_2006}, and have experimentally verified across a range of simulations and wave-analog experiments~\cite{Fromhol_phys.rev.lett_75_1142_1995, Wilkinson_nature_380_608_1996, narimanov_phys.rev.lett_80_49_1998, Honig_phys.rev.a_39_5642_1989, bogomolny_phys.rev.lett_97_254102_2006, kim_phys.rev.b_65_165317_2002, stockman_phys.rev.lett.64.2215_1990, dorr_phys.rev.lett_80_1030_1998, Sridhaar_phys.rev.lett_67_785_1991, Stein_phys.rev.lett_68_2867_1992, nockel_nature_385_45_1997, Lee_phys.rev.lett_88_033903_2002, Harayama_phys.rev.e_67_015207_2003, chinnery_phys.rev.e_53_272_1996}, culminating in the most recent direct observation of scarred eigenstates in a graphene-based quantum dot via a scanning tunneling microscopy~\cite{directVis_Ge2024}.

In general, open quantum dots (QDs) provide a versatile platform for investigating quantum scars, offering valuable insights into classical-quantum correspondence partially due to their mixed classical phase space structure. While a single scarred state is typically thought to be fragile against strong perturbations, such as system opening, the proliferation of scars can, however, result in significant robustness for the entire family of scarred states within a moderate energy window. Consequently, these scars can have a substantial effect on the transmission properties of quantum devices. For instance, non-exponential decay and tunneling anomalies attributed to scarring effects have been experimentally observed in open systems~\cite{Wilkinson_nature_380_608_1996,kim_phys.rev.b_65_165317_2002}. Similarly, theoretical studies have shown that scarred states can dominate transport in open QDs under certain conditions~\cite{scar_open_PhysRevE.77.045201}. Moreover, experimental works on mesoscopic devices have demonstrated strong energy-dependent variations in conductance, suggesting the presence of scar-mediated pathways~\cite{cabosart_nano.lett_17_1344_2017}. Notable transmission variations over small energy scales have also been attributed to the underlying chaotic dynamics of the quantum dot, which give rise to fractal conductance fluctuations~\cite{Ketzmerick_phys.rev.b_54_10841_1996, Sachrajda_phys.rev.lett_80_1948_1998}.


\begin{figure}[!t]
    \centering
    \includegraphics[width=.95\linewidth]{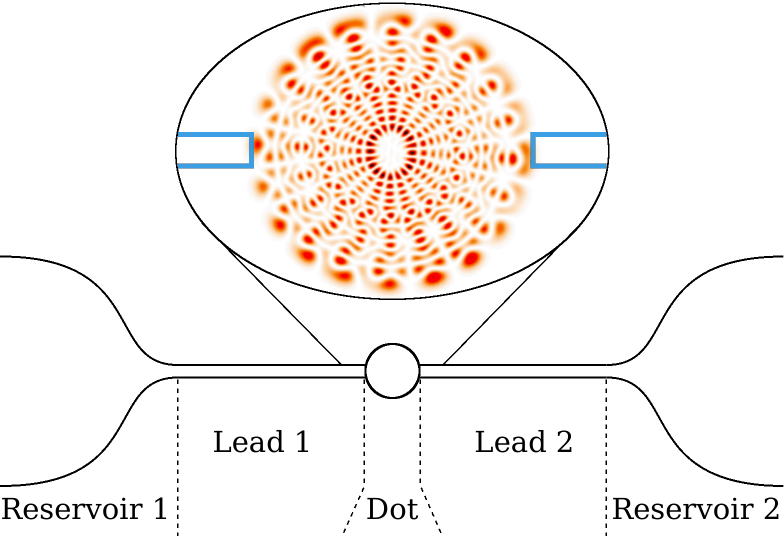}
    \caption{Schematic of the quantum transport setup: A quantum dot serves as the central scattering region, weakly coupled to two leads that are connected to electron reservoirs with different electrochemical potentials.}
    \label{fig:transSetup}
\end{figure}

In this article, we investigate the quantum transport of a two-dimensional (2D) device illustrated in Fig. \ref{fig:transSetup} in which the coupling to the reservoirs is mediated by a pair of quantum point contacts. We assume that the QD contains impurities introducing randomly positioned bumps within the confining potential, whereas a more controlled, localized disturbance is generated via a nanotip. In particular, the perturbation gives rise to a recently discovered form of scarring within the closed QD~\cite{Luukko_sci.rep_6_37656_2016, keski-rahkonen_phys.rev.b_97_094204_2017, keski-rahkonen_j.phys.conden.matter_31_105301_2019, keski-rahkonen_phys.rev.lett_123_214101_2019, Luukko_phys.rev.lett_119_203001_2017, antiscarring, simo}: these scars emerge in a synthesis of the quantum near-degeneracy in the unperturbed system and the localized nature of the perturbation. 
This unique birth mechanism~\cite{keski-rahkonen_phys.rev.lett_123_214101_2019} sets the variational scars apart from the conventional, Heller-type scarring~\cite{Heller_phys.rev.lett_53_1515_1984, kaplan_ann.phys_264_171_1998, Kaplan_nonlinearity_12_R1_1999}, extending the concept of scarring into systems lacking suitable classical POs or when those orbits are too unstable to sustain conventional scarring. In other words, this origin allows scars to emerge and thrive under conditions that would typically inhibit conventional scarring. Furthermore, it highlights the fact that the near-degeneracies of a system can have a profound impact, not only the exact degeneracies associated with the continuous symmetries, as well as underscores the subtle role that disorder can play in a quantum device. 

In particular, we demonstrate here that variational scars not only persist but also leave a distinct signature in the quantum transmission of an open QD, a fascinating prospect that has remained largely unexplored. We further support this finding by correlating the observed transmission peaks with a local density of states (LDOS)~\cite{DiVentra_book} bearing the imprint of a scar, thus paving the way towards a very first experimental observation and successive utilization of variational scarring in a QD device.


\section{System}

A common strategy to realize a QD is to apply depletion potentials to lithographically defined gates~\cite{singleElec_trans_RevModPhys.64.849, electronicQD_RevModPhys.74.1283}, which provide lateral confinement for a quasi-2D electron gas formed at the interface of a semiconductor heterostructure~\cite{fewElecQD, spinFewElec_RevModPhys.79.1217}. Theoretical and experimental studies have validated the usage of anharmonic potentials to model the external confinement of electrons in QDs~\cite{anharmonic_theo, anharmonic_2, anharmonic_Dineykhan1995,anharm_2elecQD_PhysRevB.87.155413,Halonen_Phys.Rev.B_53_6971_1996, Rasanen_Phys.Rev.B_70_115308_2004, Guclu_Phys.Rev.B_68_035304_2003}, even extending to the quantum Hall regime under strong magnetic fields~\cite{Girvin_Yang_2019,QHall_PhysRevB.52.16784, Baer2015}, with experimental evidence of scars accumulating across a wide range of systems~\cite{Fromhol_phys.rev.lett_75_1142_1995, Wilkinson_nature_380_608_1996, narimanov_phys.rev.lett_80_49_1998, Honig_phys.rev.a_39_5642_1989, bogomolny_phys.rev.lett_97_254102_2006, kim_phys.rev.b_65_165317_2002, stockman_phys.rev.lett.64.2215_1990, dorr_phys.rev.lett_80_1030_1998, Sridhaar_phys.rev.lett_67_785_1991, Stein_phys.rev.lett_68_2867_1992, nockel_nature_385_45_1997, Lee_phys.rev.lett_88_033903_2002, Harayama_phys.rev.e_67_015207_2003, chinnery_phys.rev.e_53_272_1996}. Moreover, the influence of external impurities on QDs has been quantitatively identified through differential magnetoconductance, which reveal the quantum eigenstates~\cite{Sari2022,En-nadir2023, keski-rahkonen_phys.rev.b_97_094204_2017}. 

More specifically, our device studied here constitutes a QD coupled to two electron reservoirs with distinct electrochemical potentials, serving as source and drain, mediated by a pair of quantum point contacts modeled as leads, as shown in Fig. \ref{fig:transSetup}. Throughout the discussion, all values and equations are given in (effective) atomic units (a.u.), which can be converted to SI units by considering the material parameters for, e.g., GaAs. The single-electron Hamiltonian for the closed QD device is expressed as
\begin{equation}\label{Eq:Hamiltonian}
    \bm{H}_{QD} = \frac{1}{2}\nabla^2 + V_{\text{ext}} + V_{\text{tip}} + V_{\text{imp}},
\end{equation}
where $V_{\text{ext}}=\frac{1}{2} r^2 + \frac{1}{4} r^3$ is the anharmonic oscillator external confinement potential, in which the cubic term reflects the deviations from the harmonic confinement approximation taking place at higher energies, and subsequently breaks the special degeneracy structure of harmonic oscillator. Nevertheless, due to the circular symmetry, the eigenstates of the unperturbed system are labeled by two quantum numbers \((r, m)\), corresponding to radial and angular motions, respectively. While states \((r, \pm m)\) remain exactly degenerate, the system also exhibit near-degeneracies, called resonant sets that are associated with the POs of the classical counterpart. These POs can be enumerated directly; Each PO is connected to a resonance, where the oscillation frequencies of the radial and angular motion are commensurable.

In addition to the confining potential $V_{\textrm{ext}}$, recognizing that real quantum devices are often affected by disorder arising from impurities and imperfections -- such as atoms or ions migrating into a QD system -- we model the disorder potential as a sum of randomly distributed Gaussian bumps 
\begin{equation}
    V_{\text{imp}}=A_n \sum_i \exp\left(-\frac{|\bm{r}-\bm{r}_i|^2}{2\sigma_n^2}\right),
\end{equation}
where \(A_n\) and \(\sigma_n\) represent the amplitude and width of the impurity bumps, respectively. This kind of disorder mode has been validated for a QD by density-functional theory~\cite{hirose_phys.rev.B_65_193305_2002, hirose_phys.Rev.B_63_075301_2001} and the diffusive quantum Monte-Carlo approach~\cite{Guclu_Phys.Rev.B_68_035304_2003}. Moreover, we introduce a controllable localized perturbation stemming from a conducting nanotip for which a well-established approximation~\cite{bleszynski_nano.lett_7_2559_2007, boyd_nanotechnology_22_185201_2011, blasi_phys.rev.B_87_241303_2013} is also the Gaussian profile 
\begin{equation}
    V_{\text{tip}}=A_T \exp \left(-{\frac{\vert \mathbf{r} - \mathbf{r}_0 \vert^2}{2\sigma_T^2}}\right),
\end{equation}
centered at the location \(\bm{r}_0\), where \(A_T\) and \(\sigma_T\) denote its amplitude and width. 

\section{Variational Scarring in Closed Quantum Dot}

\begin{figure*}[!t]
    \centering
    \includegraphics[width=\linewidth]{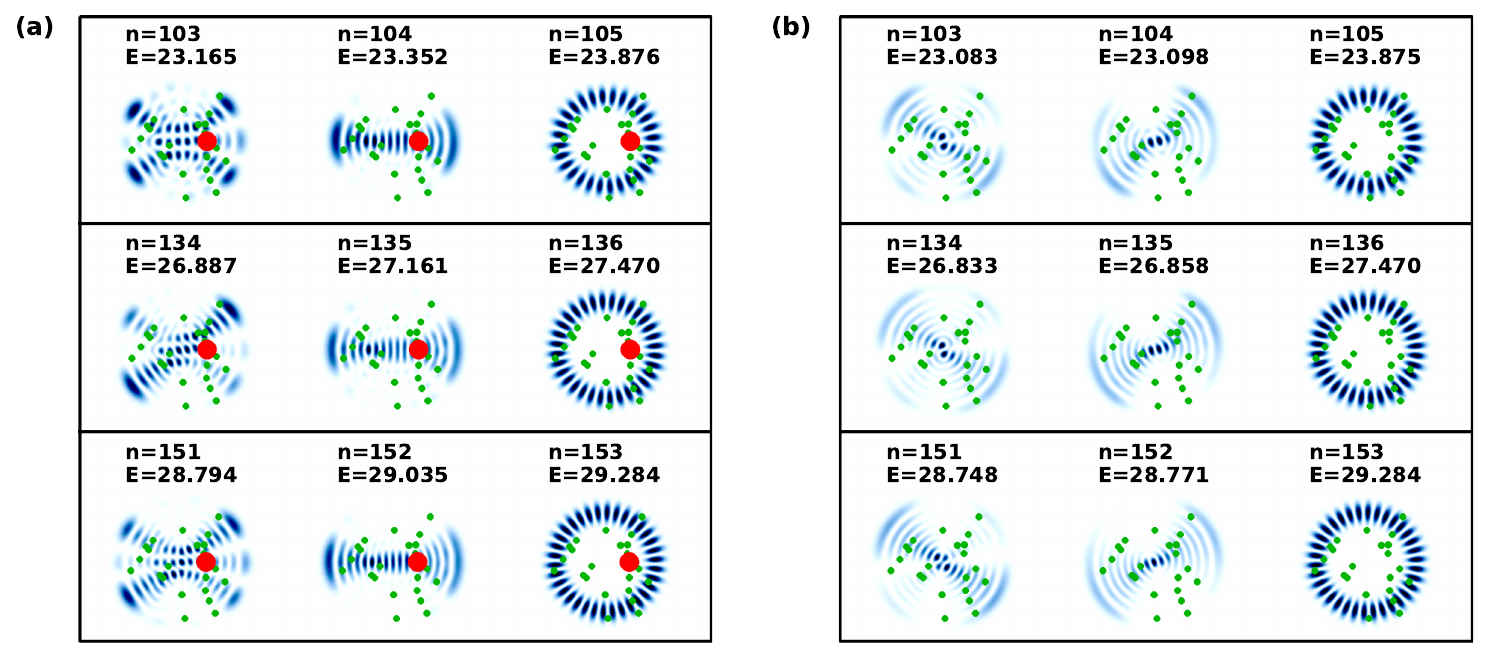}
    \caption{\textbf{(a)} Selected eigenstates of a closed, perturbed quantum dot containing both random Gaussian impurities (green dots) with average amplitude \(A_n = 2\) and width \(\sigma_n \simeq 0.05\), as well as a nanotip modeled as a Gaussian bump (red dot) with amplitude \(A_T = 12\) and width \(\sigma_T = 0.1\). These eigenstates include the first three clearly identifiable bouncing-ball scars (the middle states) and adjacent states in energy. \textbf{(b)} Corresponding eigenstates (same quantum numbers \(n\)) for the system with only random Gaussian impurities, illustrating the absence of prominent directional localization, and thereby highlighting the role of the nanotip in inducing the scarred patterns. The quantum numbers  and corresponding eigenenergies are indicated in the figure.}
    \label{fig:eigenSamples}
\end{figure*}

Before delving into the transport properties of our device, we first briefly discuss the scarring phenomenon occurring in the closed QD described above. To compute the eigenstates of the QD defined by Eq.(\ref{Eq:Hamiltonian}), we utilize the \texttt{ITP2D} software~\cite{LuukkoP.J.J.2013Itpc}, which takes advantage of the imaginary time propagation method, particularly effective for 2D systems. As predicted by the scar theory~\cite{Luukko_sci.rep_6_37656_2016, keski-rahkonen_phys.rev.lett_123_214101_2019}, a sufficient perturbation can give rise a distinct set of scarred eigenstates from a specific resonant set. Because these resonant states are linked to classical motion, certain linear combinations of these nearly-degenerate, unperturbed states create an interference pattern that traces out the path of a classical PO.  While more complex scar geometries exist\cite{Luukko_sci.rep_6_37656_2016,keski-rahkonen_phys.rev.b_97_094204_2017,keski-rahkonen_phys.rev.lett_123_214101_2019,keski-rahkonen_j.phys.conden.matter_31_105301_2019}, we focus here on the simplest but also most ubiquitous form of variational scarring where the probability density condenses along an orbit associated with the classical bouncing motion. In essence, these \emph{bouncing-ball} (BB) scars represent the quantum equivalent of radial, back-and-forth linear motion~\cite{simo}. 

As a reference, we first consider the case where there is no nanotip (i.e., $V_{tip} = 0$), and the only perturbation comes from the impurity bumps $V_{\textrm{imp}}$, with amplitude $A_n=2$ and width $\sigma_n \simeq 0.05$. In this scenario, we find that several states accommodate a BB scar. Within this impurity strength regime, the perturbation $V_{\textrm{imp}}$ mostly leads to linear combinations of a single resonant set. Therefore, we can approximate the perturbed eigenstates employing the degenerate perturbation theory, i.e., by diagonalizing $V_{\textrm{imp}}$ within the near-degenerate subspace of the given resonant set. According to the variational principle, these formed states are associated with the extrema of the expectation value of the perturbation. Due to the spatially localized nature of the perturbation, the scarred states are thus favored over non-scarred states whose probability density commonly covers a much larger region of space (see, e.g., Refs.~\cite{Luukko_sci.rep_6_37656_2016, keski-rahkonen_phys.rev.lett_123_214101_2019}). 

By the previous argument the scar orientations are mostly selected by the positions of the impurities resulting the most effective externalization of the perturbation: The BB scars shown in Fig. \ref{fig:eigenSamples}(b) orient themselves to maximize or minimize the perturbation by coinciding with or avoiding as many bumps as possible, respectively. While the scars maintain a similar alignment across a broad energy range, controlling the scar direction is challenging due to the inherent randomness of the impurity perturbation. However, by introducing an additional, stronger bump (i.e., $V_{tip} \neq 0$) via the nanotip ($A_T = 12$ and $\sigma_T = 0.1$), the BB scars align with the position of the nanotip, as seen in Fig. \ref{fig:eigenSamples}(a), which provides a feasible strategy to design the orientation of the scars experimentally.

Intriguingly, variational scarring appears to persist even up to the semiclassical limit $\hbar \rightarrow 0$, owing to the survival of resonant sets, unless destroyed by decoherence. Analogously, non-ergodic BB states in stadium billiards are proven to linger in this limit~\cite{OConnor_phys.rev.lett_61_2288_1988}. However, it remains as an open problem whether variational scars form a zero-measure subset in the semiclassical limit, as is the case for their Heller-type cousins, ergo aligning with the ergodicity theorems of Refs.~\cite{Shnirelman_Uspekhi.Mat.Nauk_29_181_1974, Colindeverdiere_comm.math.phys_102_497_1985, zelditch_duke.math.j_55_919_1987}. Moreover, this onset of ergodicity can be exceedingly slow, as exemplified by the study of a particle in a tilted box~\cite{kaplan_physica_d_121_1_1998}.

At the same time, the total perturbation responsible for scarring must be strong enough to mix eigenstates within a resonant set, yet weak enough to avoid coupling these states to the broader spectrum, which would otherwise result in chaotic-looking combinations~\cite{keski-rahkonen_j.phys.conden.matter_31_105301_2019}. As $\hbar \rightarrow 0$, the energy levels may become denser, and the resonant set may thus lose its isolation. Consequently, the same perturbation strength begins to mix a wide range of unperturbed states, thereby suppressing scarring in the high-energy regime.

Nonetheless, variational scars are not rare phenomena; In fact, they occur thrive across various disordered nanostructures. For instance, scars have been demonstrated in a broad range of power-law confinement potentials $V_{\textrm{ext}} \propto r^{a}$ ($a \in \mathbb{R}$)~\cite{Luukko_sci.rep_6_37656_2016, keski-rahkonen_j.phys.conden.matter_31_105301_2019}, as well as in a Rydberg atom set-up~\cite{Luukko_phys.rev.lett_119_203001_2017}. While homogeneousness of a confinement simplify the classical POs structure, variational scars emerge in many other types of potentials landscapes, such as non-homogeneous hyperbolic cosine wells~\cite{Luukko_sci.rep_6_37656_2016, keski-rahkonen_j.phys.conden.matter_31_105301_2019} or anisotropic oscillators~\cite{keski-rahkonen_phys.rev.lett_123_214101_2019}. A necessary element is the presence of a resonant set associated with classical POs that local perturbations can effectively mix up to form scars. Notable, exact symmetry is not essential; near-symmetries are sufficient. Indeed, scars have been demonstrated~\cite{keski-rahkonen_j.phys.conden.matter_31_105301_2019, keski-rahkonen_phys.rev.lett_123_214101_2019} in power-law potentials with non-integer exponents $\alpha$, and in elliptical oscillators or power potentials subjected to perpendicular magnetic fields. In all aforementioned cases, closed classical orbits are absent.      

On the other hand, the variational mechanism~\cite{keski-rahkonen_phys.rev.lett_123_214101_2019} is agnostic to whether the perturbative bumps are attractive or repulsive. While our discussion focus on repulsive perturbations, scars also arising from attractive bumps, i.e., potential dips, that has also been confirmed by prior studies~\cite{keski-rahkonen_phys.rev.lett_123_214101_2019, simo}. Generally, lower bump density allows the eigenstates to reflect more of the symmetry of unperturbed system, whereas higher density leads to delocalization and the disappearance of scarring. Similar effects arise from variations in bump amplitude and width. The specific behavior depends on multiple factors~\cite{keski-rahkonen_j.phys.conden.matter_31_105301_2019, keski-rahkonen_phys.rev.lett_123_214101_2019}: bump parameters (density, amplitude, width), the degree of near-degeneracy in the unperturbed spectrum, and external influences such as applied magnetic fields, hence providing several avenues for controlling scarring.

Furthermore, even in the absence of impurities, a single localized perturbation, such as from a nanotip, can be sufficient to induce distinct scars~\cite{keski-rahkonen_phys.rev.b_97_094204_2017}. However, as shown in Ref.~\cite{keski-rahkonen_phys.rev.b_97_094204_2017}, the exact shape of the scar is then highly sensitive to the tip location, while the orientation and geometry of scars caused by multiple perturbations remain robust~\cite{Luukko_sci.rep_6_37656_2016, keski-rahkonen_j.phys.conden.matter_31_105301_2019}. For example, pinned pentagram scars adjust their self-crossing point in response to radial movement of the tip, whereas BB scars, which are the primary center of attention here, exhibit robustness but can be reoriented with the tip, as also illustrated in Ref.~\cite{simo} (see also Fig.~\ref{fig:eigenSamples}).

\section{Quantum Transport Methodology }

Previous work~\cite{Luukko_sci.rep_6_37656_2016} has demonstrated variational scars to impact the transport characteristics of a QD by enabling wavepackets to propagate through the perturbed system with higher fidelity than in the corresponding unperturbed system. This boost in fidelity stems from the fact that there can be multiple strong scars oriented similarly within a moderate energy window.~\cite{Luukko_sci.rep_6_37656_2016} Moreover, these scars are robust~\cite{keski-rahkonen_j.phys.conden.matter_31_105301_2019, keski-rahkonen_phys.rev.lett_123_214101_2019, simo}, further underscoring their potential in a transport tool over the conventional scarring~\cite{Heller_phys.rev.lett_53_1515_1984}.

Nevertheless, the influence of these scars on the electrical transmission of an \emph{open} quantum device has remained elusive. To address this question, we couple the QD to two leads, effectively opening the system. All transport and related quantities are numerically computed utilizing the \texttt{TINIE} software package~\cite{tinie}, which we briefly outline below.

The leads are modeled as semi-infinite waveguides, with harmonic confinement in the transverse direction and free propagation along the longitudinal axis. This setup ensures smooth continuity and consistent boundary conditions at the interface with the scattering region.
Within the Landauer–Büttiker framework~\cite{Stefanucci_book,DiVentra_book,Datta_book}, the influence of the given lead \(\alpha\) on the central region is encapsulated in the retarded and advanced self-energies, i.e., \(\bm{\Sigma}_\alpha^R(\omega)\) and \(\bm{\Sigma}_\alpha^A(\omega)\), respectively:
\begin{align}
    \bm{\Sigma}^{R}(\omega) = & \bm{V}^{\dagger}_\alpha \; [(\omega + i\eta \mathbbm{1}) - \bm{H}_{L_\alpha}]^{-1} \; \bm{V}_\alpha \; , \\[6pt]
    \bm{\Sigma}^{A}(\omega) = & \bm{V}^{\dagger}_\alpha \; [(\omega - i\eta \mathbbm{1}) - \bm{H}_{L_\alpha}]^{-1} \; \bm{V}_\alpha \;,
\end{align}
where \(\bm{H}_{L_\alpha}\) is the isolated lead \(\alpha\) Hamiltonian, \(\mathbbm 1\) is the identity matrix in the lead subspace, and \(\eta>0\) is an infinitesimal parameter that fixes the boundary conditions.  

The matrix \(\bm{V}_\alpha\) describes the coupling between the lead–dot interface slice of lead \(\alpha\) and the adjacent sites in QD. It is constructed by projecting the full device Hamiltonian \(\bm{H}_{\text{tot}}\) onto the basis functions of the lead and the dot. In block form:
\begin{equation}
\bm{H}_{\text{tot}}=
\begin{pmatrix}
\bm{H}_{L_1}          & \bm{V}_{L_1}           & 0                \\
\bm{V}_{L_1}^{\dagger} & \bm{H}_{\text{QD}} & \bm{V}_{L_2}^{\dagger}\\
0                 & \bm{V}_{L_2}           & \bm{H}_{L_2}
\end{pmatrix},
\label{eq:Htot}
\end{equation}
where \(\bm{H}_{\text{QD}}\) is the isolated-dot Hamiltonian, \(\bm{H}_{L_{1/2}}\) are the lead Hamiltonians, and
\(\bm{V}_{1/2}\) are the lead–dot coupling blocks.  Accordingly, the elements of \(\bm{V}_\alpha\) are
\begin{equation}
    [\bm{V}_\alpha]_{ij} = \langle \phi_{\alpha,i} | \bm{H}_{\text{tot}}| \psi_{\text{QD},j} \rangle,
\end{equation}
with \(\phi_{\alpha,i}\) a basis function on the interface slice of the lead \(\alpha\), and \(\psi_{\text{QD},j}\) a basis function in the dot region.

The coupling strength of the lead–dot contact is quantified by the energy-dependent total rate operator \(\bm{\Gamma}_{\text{tot}}(\omega)=\sum_\alpha \bm{\Gamma}_\alpha(\omega)\) with the lead-specific rates 
\begin{equation}
    \bm{\Gamma}_\alpha = i[\bm{\Sigma}_\alpha^R - \bm{\Sigma}_\alpha^A],
\end{equation}
reflecting the level broadening induced by coupling to an individual lead \(\alpha\). Since \(\textrm{Tr}[\bm{\Gamma}_{\text{tot}}(\omega)]\) represents the total broadening of QD energy levels due to coupling with the leads, we define a dimensionless coupling parameter as
\begin{equation}
    \mathcal{C} = \textrm{Tr}[\bm{\Gamma}_{\text{tot}}(\omega)] / \delta
\end{equation}
where \(\delta\) is the mean level spacing of the isolated QD within the energy window of interest. This ratio distinguishes between weak coupling (\(\mathcal{C}\ll1\)), where transmission is dominated by sharp, isolated resonances, and intermediate to strong coupling (\(\mathcal{C}\gtrsim1\)), where significant level broadening leads to overlapping resonances and more delocalized transport pathways. 

The transmission between two distinct leads \(\alpha\) and \(\beta\), through the QD is then determined as
\begin{equation}
    \mathcal{T}_{\alpha \beta}(\omega) = \textrm{Tr}\Big[\bm{G}^R(\omega) \, \bm{\Gamma}_\beta(\omega) \, \bm{G}^A(\omega) \, \bm{\Gamma}_\alpha(\omega)\Big],
    \label{eq:transmission}
\end{equation}
where $\hat{G}^R(\omega)$ and $\hat{G}^A(\omega)$ denote the retarded and advanced Green's functions of the open quantum system, respectively~\cite{Stefanucci_book}. These Green's functions incorporate the effect of coupling to the leads through the retarded self-energies $\Sigma_\alpha^R(\omega)$, and are defined as
\begin{align}
\bm{G}^R(\omega) &= \left[\omega\mathbbm{1} - \bm{H}_{\text{QD}} - \sum_{\alpha}\Sigma_\alpha^R(\omega)\right]^{-1} = [\bm{G}^A(\omega)]^\dagger.
\end{align}

In addition to transmission coefficient of Eq.~\ref{eq:transmission}, we compute LDOS to resolve the spatial distribution of quantum states contributing to transport at specific probe energies \(\omega\). As defined in the context of Green’s function formalism~\cite{Datta_book}, the LDOS is given by
\begin{equation}
    \rho(\bm{r},\omega) = -\textrm{Im} \big[\bm{G}^R(\bm{r},\omega)\big], 
\end{equation}
where \(\bm{G}^R(\bm{r},\omega)\) is subsequently the projection of the retarded Green's function in real space~\cite{DiVentra_book}, evaluated at position \(\bm{r}\) as
\begin{equation}
    \bm{G}^R(\bm{r},\omega) = \sum_{i,j} \psi_{\text{QD},i}^*(\bm{r}) \, [\bm{G}^R(\omega)]_{ij} \, \psi_{\text{QD},j}(\bm{r}).
\end{equation}
This measure serves as a powerful diagnostic, opening up a way to correlate transport features of the open system with the underlying scarred wavefunction structure of the closed counterpart.

\section{Transport Signatures of Variational Scars}
\begin{figure}[!t]
    \centering
    \includegraphics[width=\linewidth]{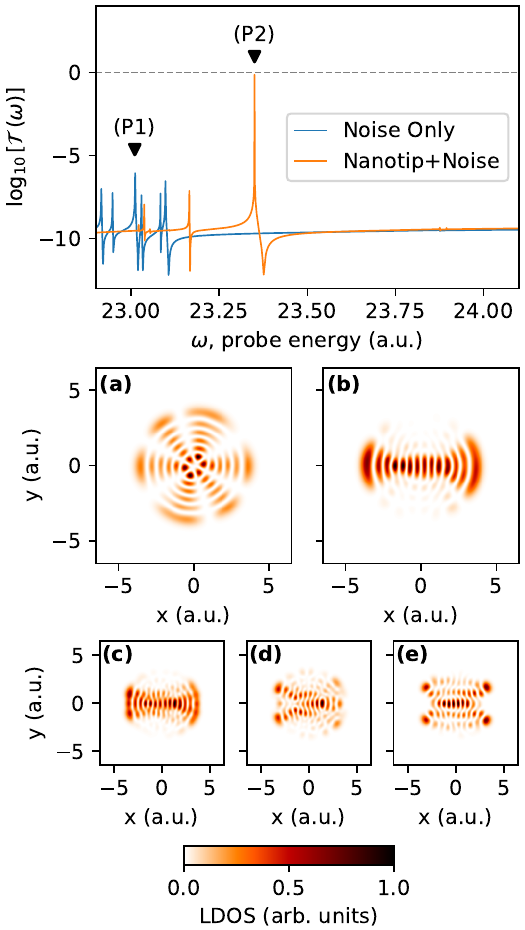}
    \caption{Logarithm of transmission vs. probe energy within bias window \(23–24\) for noise-only case with \(\mathcal{C}\simeq 0.001\) (blue) and nanotip+noise with \(\mathcal{C}\simeq 0.002\) (orange), both in weak coupling limit (\(\mathcal{C}\ll 1\)). The strongest peaks are labeled as (P1) for noise-only and (P2) for nanotip+noise. Logarithmic scale is used due to the wide range of transmission values. Below, normalized LDOS maps are shown at the corresponding peak energies (P1) and (P2) in panels (a) and (b), respectively, illustrating distinct bouncing-ball scar pattern induced by the tip. Panels (c)-(e) display normalized LDOS of nanotip+noise system at the energy corresponding to strongest transmission peak in weakly coupled leads (\(\omega=23.350\)), now under stronger coupling strengths \(\mathcal{C}\): (c) \(\simeq 7.55\), (d) \(\simeq 48.34\), and (e) \(\simeq 121.20\).}
    \label{fig:weak-coupling}
\end{figure}

\begin{figure}[!t]
    \centering
    \includegraphics[width=\linewidth]{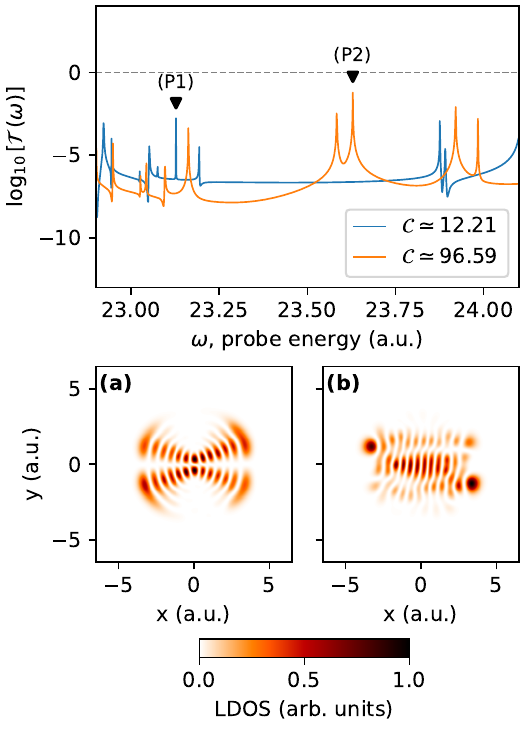}
    \caption{Logarithm of transmission vs. probe energy within bias window \(23–24\), shown for intermediate (\(\mathcal{C}\simeq 12.21\), blue), and strong (\(\mathcal{C}\simeq 96.59\), orange) coupling strengths. Prominent transmissions labeled as (P1) at  \(\omega=23.127\) and (P2) at energy \(\omega=23.629\). Below, the corresponding normalized LDOS maps at these peak energies are shown in panels (a) and (b), respectively, revealing how increased self-energy disrupts the scarred patterns observed at weaker coupling, resulting in more extended wavefunction distributions. 
}
    \label{fig:strong-coupling}
\end{figure}
In our transport setup, we choose to posit the tip strategically so that the induced BB scars oriented in conjuncture to the leads and thus opening a direct pathway through the QD. In particular, we identify a prominent BB scarred state at energy 23.352 in the close system when exposed to the nanotip. To quantify its impact on electron transport, we focus on a bias voltage range of \(23-24\) and subsequently assess \(\mathcal{T}(\omega)\) and \(\rho(\bm{r},\omega)\) with an energy resolution of \(d\omega=10^{-3}\) within this interval.


Figure \ref{fig:weak-coupling} shows the transmission $\mathcal{T}(\omega)$ of the device as a function of energy $\omega$ in the weak coupling regime, i.e., \(\mathcal{C}\ll 1\). As in the scarring analysis above, we compare two cases: one with only impurity disorder $V_{\textrm{imp}}$ and coupling strength \(\mathcal{C}\simeq 0.001\), and another with a nanotip also present, $V_{\textrm{imp}}+V_{\textrm{tip}}$, and \(\mathcal{C}\simeq 0.002\). Notably, we observe significantly higher transmission peaks in the nanotip scenario [peak labeled (a) in Fig. \ref{fig:weak-coupling}] compared to the noise-only instance [peak labeled (b) in Fig. \ref{fig:weak-coupling}]. We attribute this boost in transmission to the presence of BB scarring, where the noise-only case lacks the coherent pathways formed by the lead-oriented scars. This interpretation is confirmed by the LDOS, which reveals a pattern at the transmission peak energy [Fig. \ref{fig:weak-coupling}(a)] corresponding to a BB scar of the closed QD, a feature absent in the nanotip-free benchmark [Fig. \ref{fig:weak-coupling}(b)]. 

To further investigate the role of the nanotip, we performed additional simulations where the tip position was systematically varied along the axis connecting the two leads. Our results indicate that while the nanotip consistently enhanced the transmission relative to the noise-only scenario, the specific configuration selected for our primary results exhibited the strongest amplification. This optimal positioning clearly illustrates the significant impact of nanotip-induced BB scarring on electron transport.

On the other hand, panels (c)-(e) of Fig. \ref{fig:weak-coupling} present the LDOS for the nanotip case at the same energy of 23.350, corresponding to the strongest transmission peak in the weak coupling limit. These cases are associated with three different coupling strengths, \(\mathcal{C}\simeq 7.55,  48.34,  121.20\), respectively, from intermediate to strong coupling limit. As the coupling strength increases, the LDOS pattern that initially exhibits a clear scar becomes gradually smeared and chaotic. This reflects the fact that stronger coupling results in greater self-energy, which in turn broadens QD's energy levels. The resulting spectral broadening diminishes the influence of individual scarred states on electron transport by delocalizing the transmission pathways. 

To further explore the stronger coupling limit (\(\mathcal{C}\gtrsim1\)), we focus on the same energy range where the nanotip-induced transport enhancements occur under weak coupling. Therefore, Fig. \ref{fig:strong-coupling} shows the transmission for two cases of intermediate and strong couplings ($\mathcal{C} \simeq 12.21$ and $\mathcal{C} \simeq 96.59$). As illustrated in Fig. \ref{fig:strong-coupling}(a) and Fig. \ref{fig:strong-coupling}(b), the level broadening caused by the leads disrupts the scar patterns seen in the weak coupling regime, replacing them with more extended density distributions. Furthermore, this pronounced level broadening alters the line shape of the transmission peaks. In the weak coupling regime, the QD exhibits asymmetrical, Fano-like resonances~\cite{Fano_PhysRev.124.1866,Fano_nano_RevModPhys.82.2257} in the transmission; whereas, in the stronger coupling case, the increased broadening of the QD energy levels smooths these resonances, resulting in more symmetric transmission peak profiles.


An intriguing direction for exploration is the use of a nanotip to orient the scars perpendicular to the leads. In this configuration, the scars would couple less effectively to the leads, allowing them to persist even in the strong coupling regime. However, this setup would invert the current situation, with the scars now associated with transmission dips rather than peaks. This effect could be interpreted as a consequence of antiscarring~\cite{antiscarring}, leading to a reduced decay rate in the open system~\cite{kaplan_phys.rev.e_59_5325_1999}. Furthermore, we anticipate that applying an external magnetic field~\cite{keski-rahkonen_phys.rev.b_97_094204_2017} could offer additional control over scarring, potentially enabling scar-driven transport in more complex architectures, such as three-terminal devices.

On the experimental front, our study lays a foundation for further exploration in QD-based devices, where gate-defined potentials can replicate customized confinement, and precise impurity positioning can simulate nanotip perturbations. Inspired by recent advancements in mapping scars, scanning tunneling microscopy (STM) appears to be a viable option to verify the existence of variational scarring, as suggested in Ref.~\cite{keski2024variational}, paving a way to study these scars in open systems. Parallel to the STM path, our quantum device provides a robust platform for using scanning gate microscopy (SGM) as a direct method to probe scarred states in relation to the transmission: The SGM tip~\cite{SGM_tip_PhysRevB.108.195415}, biased to create a local perturbation by depleting electron density, allows for spatially resolved conductance investigations of a QD device~\cite{SGM_imaging}.

\section{Summary}
To summarize, we investigated the influence of variational scarring on electron transport in two-dimensional quantum dots. In particular, we demonstrated that these scars not only survive but can also facilitate a controlled modulation of the transmission. In the weak coupling regime (\(\mathcal{C} \ll 1\)), scars enhance the transmission through the device, when oriented along the leads with a controlled perturbation, such as that generated by a nanotip. However, in the intermediate to strong coupling limit (\(\mathcal{C} \gtrsim 1\)), the effect of an individual scar is weakened by the non-scarred transport channels. Nevertheless, our study marks a step toward establishing a basis for further experimental exploration of this intriguing subclass of quantum scarring and harnessing its potential for nanoelectronic applications.


\section{Acknowledgements}

We acknowledge CSC—Finnish IT Center for Science for computational resources, and Research Council of Finland, ManyBody2D Project No. 349956, for financial support. J.K.-R. thanks the Oskar Huttunen Foundation for the financial support. This project is also supported by the National Science Foundation (Grant No.~2403491). 

\bibliography{Bibliography}

\end{document}